\documentclass[aps,prc,twocolumn,groupedaddress,showpacs]{revtex4}

\usepackage{graphicx}

\begin{document}

\title{Effective nucleon mass and the nuclear caloric curve}
\author{D.V. Shetty}
\affiliation{Cyclotron Institute, Texas A$\&$M University, College Station, TX 77843, USA}
\affiliation{Physics Department, Western Michigan University, Kalamazoo, MI 49008, USA}
\author{G.A. Souliotis, S. Galanopoulos, S.J. Yennello}
\affiliation{Cyclotron Institute, Texas A$\&$M University, College Station, TX 77843, USA}
\date{\today}

\begin{abstract}
Assuming a schematic form of the nucleon effective mass as a function of nuclear excitation energy and mass, we provide a simple explanation 
for understanding the experimentally observed mass dependence of the nuclear caloric curve. It is observed that the excitation energy 
at which the caloric curve enters into a plateau region, could be sensitive to the nuclear mass evolution of the effective nucleon mass. 
\end{abstract}


\maketitle

The experimentally observed ``plateau" in the nuclear ``caloric curve" (temperature versus excitation energy) has long
been seen as a signature of liquid-gas phase transition, similar to that in real fluids. Recently, Sobotka 
{\it {et al.}} \cite{SOB04, SOB06}, have shown that the plateau in the nuclear caloric curve is a consequence of a combined 
effect of decreasing density due to thermal expansion, and the evolution of in-medium nucleon effective mass, rather than 
an indication of liquid-gas phase coexistence. This model for describing the caloric curve is based on the relaxation of density 
profile of the mononucleus that results in maximum entropy under a local density approximation for the level density parameter. A 
parametric form of the base density was assumed and the entropy was calculated in the Fermi-gas model. The evolution of effective 
mass with density and excitation was included in a schematic fashion as it is currently unknown.
\par
The plateau in the caloric curve from this model is  established at a rather modest excitation energy (about 2 MeV/nucleon), 
well below the excitation energy where the experimentally observed caloric curve enters into a plateau region. Also, the value of 
the plateau temperature remains the same for a mass, $A$ = 90 system as it does for a mass, $A$ =  197 system. However, 
experimentally determined caloric curve \cite{NAT02} shows plateau temperature that decreases with increasing mass, $A$.
\par
In this work, we investigate how the mass dependence of the caloric curve, can be understood using the above concept of effective 
nucleon mass and thermal expansion in a simple phenomenological approach. In particular,  we present a schematic expression to 
understand the experimentally observed plateau in the caloric curve, and  show that the observed mass dependence of the caloric 
curve  can be reproduced  using effective nucleon mass as a function of excitation energy, that is mass dependent. 
\par
We begin with the assumption that the relation between the temperature, $T$, and the experimentally measured 
total excitation energy, $E^{*}$, for an expanding nucleus of mass $A$, can be expressed by a form analogous to that for the Fermi gas,

\begin{equation} 
                    E^* = \frac{T^2}{K_{eff}(\rho,A)}
\end{equation}

where $K_{eff}$ is the inverse nuclear level density parameter of the hot and dilute nucleus at density $\rho$, and written as,

\begin{equation} 
                    K_{eff}(\rho, A) =   \frac{4\epsilon_{F}(\rho,A)}{\pi^{2}}   
\end{equation}

where, $\epsilon_{F}(\rho,A)$, is the Fermi energy of the finite and expanding nucleus and given as,

\begin{equation} 
                    \epsilon_{F}(\rho,A) =   \epsilon_{F}^o \biggl[\frac{m^{*}(\rho_o, A)}{m^{*}(\rho, A)}\biggr] \biggl( \frac{\rho} {\rho_{o}} \biggr)^{2/3}  
\end{equation}

The quantity $m^{*}(\rho_o,A)$, is the ratio of the effective mass of the nucleon to the mass of the free nucleon, assuming the
 nucleus to be a Fermi gas and at ground state ($T$ = 0). The quantity $m^{*}(\rho,A)$ is the ratio of the effective mass of the nucleon 
to the mass of the free nucleon, in hot ($T$ $>$ 0) and expanding finite nucleus density $\rho$.
\par
From the above equations, one can write the inverse nuclear level density parameter of the hot and dilute nucleus as \cite{NOR02},

\begin{equation} 
                    K_{eff}(\rho, A) = K_{o}  \biggl(\frac{\rho}{\rho_o} \biggr)^{2/3} \biggl[\frac{m^{*}(\rho_o, A)}{m^{*}(\rho, A)}\biggr] 
\end{equation}

where, $K_{o} = 4\epsilon_{F}^o/\pi^{2}$ $\approx$ 15, is the inverse level density parameter of uniform, non$-$dissipative Fermi gas.
$\epsilon_{F}^o$, and $\rho_o$ are the Fermi energy and the nuclear saturation density at $T$ = 0 MeV.
\par
From the above two expressions, Eqs. 1 and 4, one can write the temperature versus the excitation energy for a nucleus expanding to an equilibrium 
density $\rho$,

\begin{equation} 
                 T^2 = K_{o} \biggl[ \frac{m^{*}(\rho_o, A)}{m^{*}(\rho, A)} \biggr] \biggl( \frac{\rho} {\rho_{o}} \biggr)^{2/3} E^{*}   
\end{equation}

\par

Alternatively, one can also start with the assumption that the total excitation energy $E^*$, of an expanding nucleus can be written as, 

\begin{equation}
          E^{*} = E^{*}_{ther} + E_{exp}
\end{equation}

where, $E^{*}_{ther} = T^{2}/K_{o}[m^{*}(\rho_{o}, A)/m^{*}(\rho, A)]$ is the thermal part of the excitation energy, and $E_{exp} = \epsilon _{b}(1 - \rho /
\rho_{o})^2$ is the expansion energy of the finite nucleus. The expansion energy assumed in the above expression is a simple upside down bell 
shaped, suggested by Friedman \cite{FRI88}, with the ground state binding energy, $\epsilon _{b}$ = 8 MeV. The temperature versus the 
total excitation energy relation for a nucleus expanding to an equilibrium density, $\rho$, can then be written as,

\begin{equation}
                  T^{2} = K_{o}\frac{m^{*}(\rho_{o}, A)}{m^{*}(\rho, A)}[E^* - \epsilon_{b}(1 - \rho / \rho_{o})^2]
\end{equation}

Using equations, 5 and 7, we can now study the experimental caloric curve and its mass dependence.
\par
Fig. 1 shows the experimentally measured caloric curve for the mass range of $A$ = 100 - 140, from various measurements 
compiled by Natowitz {\it {et al.}} \cite{NAT02}.  The data from all different measurements are shown collectively in the figure 
by inverted triangle symbols and no distinction is made between them. The dotted (black) curve in the figure is the simple Fermi 
gas relation, $E_{ther}^* = T^{2}/K_{o}$, with the inverse level density parameter, $K_{o}$ = 15. 
The results of the equations 5 and 7 are shown by the solid (blue) and dashed (red) curves, respectively. In both equations, the 
density $\rho / \rho_{o}$, for a given excitation energy $E^{*}$, was taken to be that 
adopted by Bondorf {\it {et al.}} \cite{BON85, BON98} (shown by the solid black curve in Fig. 2(b)). 
    \begin{figure}
    \includegraphics[width=0.48\textwidth,height=0.3\textheight]{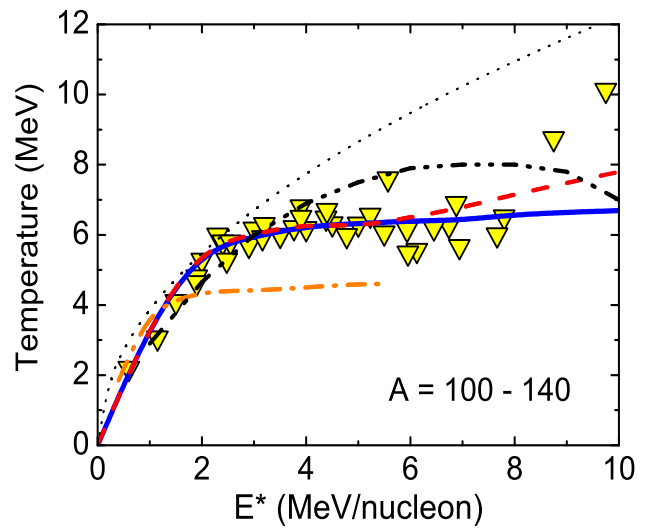} 
    \caption{(Color online) Temperature as a function of excitation energy for mass $A$ = 100 - 140. The data points 
    (inverted triangles) are from Ref. \cite{NAT02}. The dotted curve is the Fermi gas relation. The dot-dot-dashed curve is from Ref. \cite{DE06}. 
The dot-dashed curve is from Ref. \cite{SOB06}. The solid and the dashed curve are from Eq. 5 and 7, respectively. }
    \end{figure}
    \begin{figure}
    \includegraphics[width=0.45\textwidth,height=0.5\textheight]{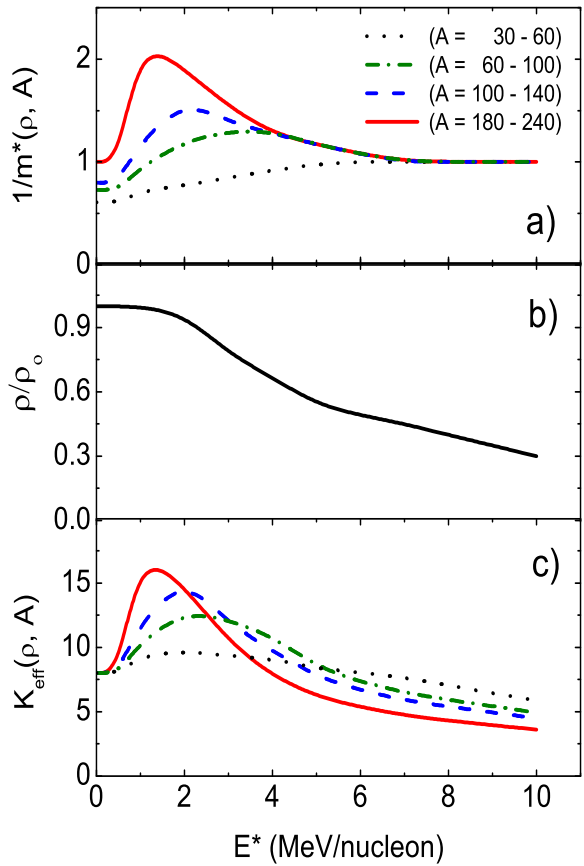} 
    \caption{(Color online) a) Effective nucleon mass ratio, b) Density and, c) $K_{eff}$, as a function of excitation energy. The different
    curves in a) and c) correspond to the different mass ranges.}
    \end{figure}
The effective mass $m^{*}(\rho, A)$, as a function of excitation energy was assumed to have an empirical dependence of the form shown by the 
dashed (blue) curve in Fig. 2(a).  From Fig. 1, one observes that Eq. 5 (solid, blue 
curve) and Eq. 7 (dashed, red curve) with the same effective mass and density dependence of the excitation energy gives similar results up to 
excitation energy of 6 MeV/nucleon, with a small difference at higher excitation energies. The difference at higher energies is due to the 
different form of the expansion energy assumed in the two expressions. Both calculations, show a plateau at excitation energy 
above 3 MeV/nucleon, in good agreement with the experimental data.
    \begin{figure}
    \includegraphics[width=0.45\textwidth,height=0.67\textheight]{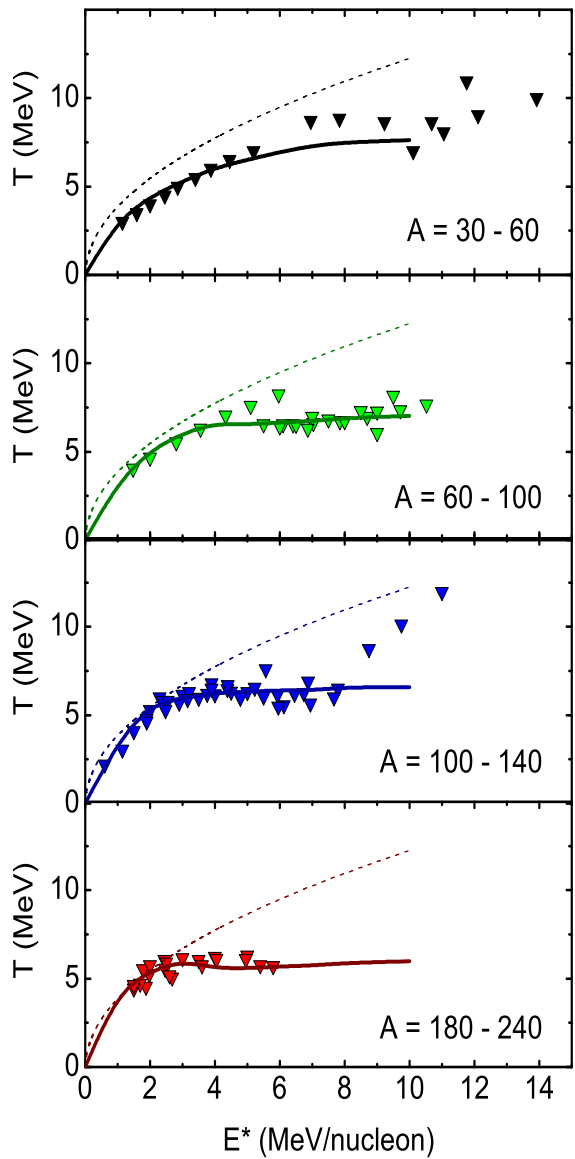} 
    \caption{(Color online) Temperature as a function of excitation energy for various mass ranges. The data points 
    (inverted triangles) are from Ref. \cite{NAT02}. The solid curves are obtained from Eq. 5. The dashed curve is the
    Fermi gas relation.}
    \end{figure}
    \begin{figure}
    \includegraphics[width=0.45\textwidth,height=0.67\textheight]{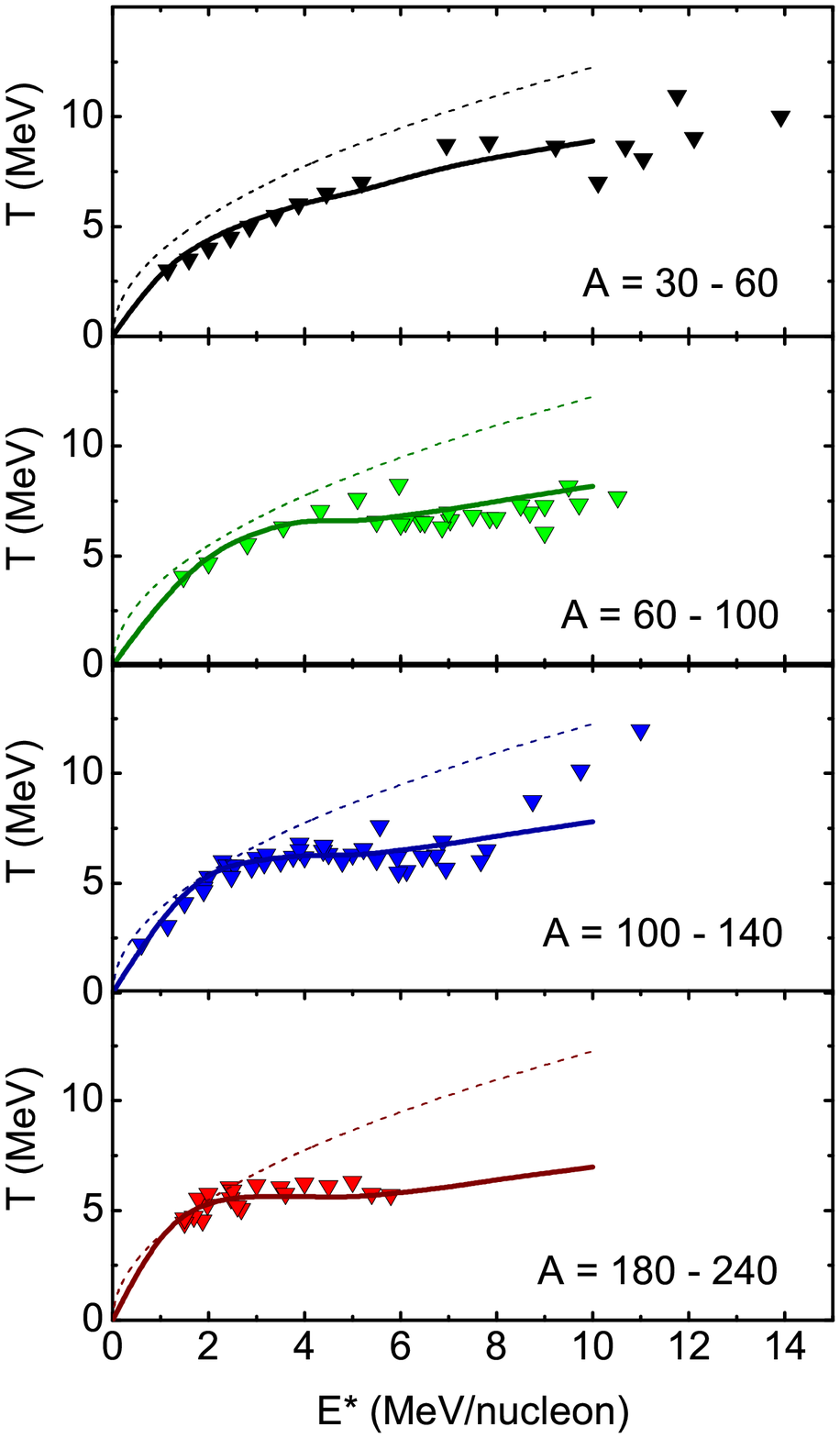} 
    \caption{(Color online) Temperature as a function of excitation energy for various mass ranges. The data points 
    (inverted triangles) are from Ref. \cite{NAT02}. The solid curves are obtained from Eq. 7. The dashed curve is the
    Fermi gas relation.}
    \end{figure}

\begin{table}
\caption{\label{tab:table1} The ground state effective mass, $m^{*}(\rho_o, A)$, used in Eq. 3 and 5,
for various nuclear mass range.}
\begin{ruledtabular}
\begin{tabular}{ccccc}
 &  Mass ($A$)   & $m^{*}(\rho_o, A)$    \\
\hline                                         
 &  30 - 60    &   0.87                \\
 &  60 - 100   &   0.73                \\
 &  100 - 140  &   0.67                \\   
 &  180 - 240  &   0.53                \\   
\end{tabular}
\end{ruledtabular}
\end{table}

    \begin{figure}
    \includegraphics[width=0.47\textwidth,height=0.3\textheight]{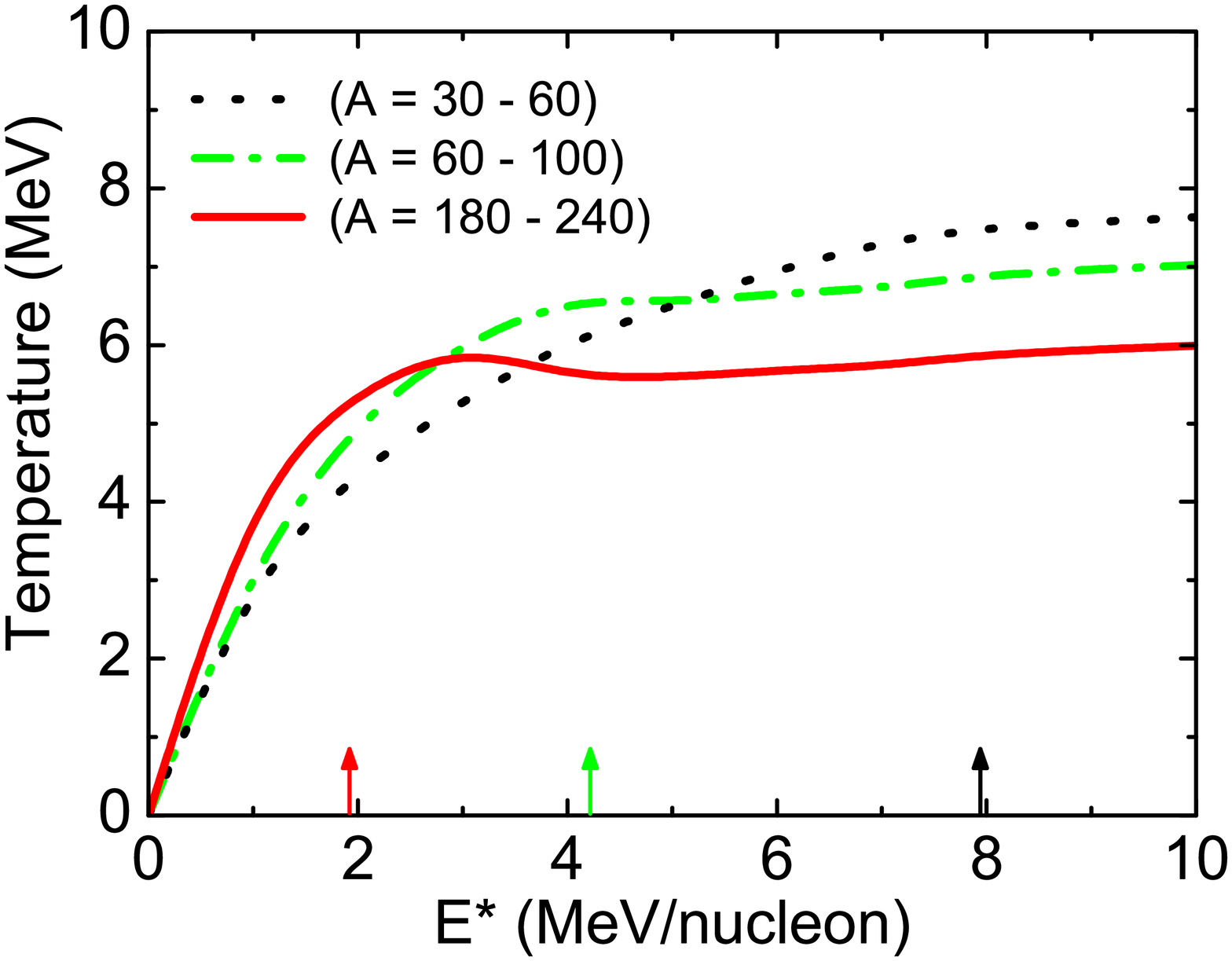} 
    \caption{(Color online) Comparison between the caloric curves obtained from Eq. 5 for three different mass ranges.}
    \end{figure}
The caloric curve obtained from Eq. 5 and 7 is in much better agreement with the data compared to those obtained by Sobotka {\it {et al.}} 
\cite{SOB06}. The result of Sobotka {\it {et al.,}} is shown by the dot-dashed (orange) curve. Also shown in the figure is the result of 
De {\it {et al.}} \cite{DE06}, for $A$ = 150 (dot-dot-dashed, black curve). The calculation of De {\it {et al.}} uses a realistic effective Hamiltonian 
to calculate the base density profile in a Thomas-Fermi framework with the entropy calculated microscopically. The above comparison shows that 
the phenomenological expressions 5 and 7 can be used to understand the characteristic features of the caloric curve.
\par
In the following we  use Eq. 5 and 7, to study the mass dependence of the caloric curve. Fig. 3 and 4 shows the experimental caloric curve data 
(inverted triangles) for the different mass ranges obtained from the work of Natowitz {\it {et al.}} \cite{NAT02}. The results of Eq. 5 and 7 are 
shown by the solid curves in Fig. 3 and 4, respectively. The empirical form of the effective nucleon mass and the density as a function of 
excitation energy used in Eq. 5 and 7 for different  mass regions are as shown in Fig. 2(a) and 2(b). These were obtained by tuning the data with a fixed 
$K_{o}$ parameter. The effective mass for the ground state 
nucleus $m^*(\rho_o, A)$, is as shown in Table I.  The choice of the $m^{*}(\rho_o, A)$ value was dictated by its sensitivity to the temperature 
at which the caloric curve reaches the plateau region. This is discussed in the following paragraph. The Fermi gas caloric curve is shown by the 
dotted curves in Fig. 3 and 4.
\par
An important point to note in Fig. 3 and 4 is the evolution of the excitation energy of the entry point into the caloric curve plateau 
with nuclear mass. To illustrate this point more clearly, we show in Fig. 5, a comparison between the caloric curves for $A$ = 30 - 60, $A$
= 60 - 100, and $A$ = 180 - 240 mass regions, obtained using Eq. 5. The arrows in the figure correspond to the approximate values of the excitation energy 
at which the caloric curve for each of the three masses enter into a plateau region. The figure shows that the temperature at which the plateau is 
reached is sensitive to the value of the $m^*(\rho_o, A)$. A higher $m^*(\rho_o, A)$ results in higher plateau temperature for lighter mass. The 
excitation energy 
at which the plateau is reached is, on the other hand, sensitive to the mass dependence of the effective nucleon mass, $m^*(\rho, A)$. This excitation energy 
corresponds to the energy at which the 1/$m^{*}(\rho, A)$, shown in Fig. 2(a), peaks. The shift in the peak to higher excitation energy for 
decreasing mass of the system results in a plateau being reached at higher excitation energy for lighter mass. To explain the experimentally 
observed caloric curve, effective nucleon mass ratio that is dependent on the excitation energy$/$density and the nuclear mass therefore seems imperative. 
The effective nucleon mass as a function of excitation energy for different masses, shown in Fig. 2(a), and used in the above analysis, is an empirical deduction. 
At present, we do not know of any formal approach to deduce such a dependence. Qualitatively such a dependence is expected in the interior of the nucleus, 
where the effective mass is reduced in the bulk, peaks at the surface and reduces to one with decreasing density and increasing excitation energy \cite{HAS86}. 
In such a prescription, the effective nucelon mass is often given by phenomenological expressions that includes a momentum dependent and a frequency dependent 
term. The detailed density and excitation energy dependence of these terms are unknown. Theoretical investigation in this direction would therefore be interesting. While a formal understanding of the effective nucleon mass as a function of excitation energy and density over a range of nuclear mass would  be very fruitful, the emperical approach utilizing the caloric curve in this work can aid in studying the isospin ($N/Z$) dependence of the effective mass in asymmetric nuclei. 
\par
In the following we show that the mass dependence of the inverse level density parameter $K_{eff}$, obtained using the above  empiricaly deduced 
effective nucleon mass, $m^*(\rho, A)$, is consistent with the experimentally and theoretically deduced inverse level density parameter at low exciation energy. In the past, temperature dependence of the nuclear level density parameter has been investigated extensively by studying 
the spectra of light particles emitted in hot nuclei populated at $E^{*}$ $>$ 1 MeV/nucleon ($T$ $>$ 2 MeV). It has been shown from these studies 
\cite{NEB86, HAG88} that the inverse level density parameter $K$, increases from 8 to 13 for temperature increasing from $T$ = 2 MeV to 
$T$ = 5.5 MeV in nuclei of mass $A$ $\sim$ 160. However, similar studies \cite{CHB91, FOR91, YOS92} carried out for light nuclei, such 
as $A$ $\sim$ 40, failed to show an increase. The $K$ remaining nearly constant at 9 - 10 in the excitation energy range of 2.5 to 5.0 MeV/nucleon. 
In Fig. 2(c), we show the effective inverse level density parameter, $K_{eff}$, of the hot and dilute nucleus for various mass regions obtained 
from the present analysis. For $T$ = 0 MeV, one observes that the inverse level density parameter, $K_{eff}$ $\sim$ 8, in agreement with the low excitation 
energy studies. For light nuclei, $A$ = 30 - 60, the $K_{eff}$ remains essentially constant up to excitation energy of 5 MeV/nucleon. While for heavier nuclei 
it varies between 8 and 16 for excitation energies below 2 - 3 MeV. Similar dependence was also obtained from the theoretical calculations of 
Ref. \cite{SHL91, DE98} that investigated the mass dependence of the level density parameter at low exciation energy. For higher excitation energies, 
Fig. 2(c) shows a steady decline in the inverse level density parameter, which results in a plateau like behavior in the caloric curve (temperature versus 
excitation energy plot) for the hot and expanding nucleus.  
\par
In conclusion, it is shown that the mass dependence of the nuclear caloric curve can be modelled  using Equations 5 and 7, with the effective nucleon mass
of the form shown in Fig. 2(a), or alternatively, using Eq. 1 with the inverse level density parameter of the form shown in Fig. 2(c). The mass dependence of the 
inverse level density parameter thereby obtained is consistent with the experimentally and theoretically deduced level density parameter 
for low excitation energies. Furthermore, the present study demonstrates that the caloric curve can be used as a tool to fine tune the effective nucleon mass and study the nuclear interaction. A natural extension of the present study would be to apply the present approach to fine tune the effective mass further by including the  isospin ($N/Z$) dependence. Such a study would require caloric curve measurements of asymmetric ($N/Z$ $>$ 1) nuclei using beams of radioactive nuclei. This would provide a complete understanding of the interaction under extreme conditions of excitation energy, density and isospin ($N/Z$).
\par
This work was supported in part by the Robert A. Welch Foundation through grant No. A-1266, and the Department of Energy through grant No. DE-FG03-93ER40773.


\begin{thebibliography}{}

\bibitem{SOB04} L.G. Sobotka, R.J. Charity, J. Toke, and W.U. Schroder, Phys. Rev. Lett. {\bf 93}, 132702 (2004).
\bibitem{SOB06} L.G. Sobotka and R.J. Charity, Phys. Rev. C {\bf 73}, 014609 (2006).
\bibitem{NAT02} J.B. Natowitz, R. Wada, K. Hagel, T. Keutgen, M. Murray, A. Makeev, L. Qin, P. Smith, and C. Hamilton, Phys. Rev. C {\bf 65}, 034618 (2002). 
\bibitem{NOR02} W. Norenberg, G. Papp, and P. Rozmej, Eur. Phys. J. A{\bf 14}, 43 (2002).
\bibitem{FRI88} W.A. Friedman, Phys. Rev. Lett. {\bf 60}, 2125 (1988). 
\bibitem{BON85} J.P. Bondorf, R. Donangelo, I.N. Mishustin, and H. Schulz , Nucl. Phys. A{\bf 444}, 460 (1985).
\bibitem{BON98} J.P. Bondorf, A.S. Botvina, and I.N. Mishustin, Phys. Rev. C {\bf 58}, 27 (1998). 

\bibitem{DE06} J.N. De, S.K. Samaddar, X. Vinas, and M. Centelles, Phys. Lett. B{\bf 638}, 160 (2006).
\bibitem{HAS86} R.W. Hasse and P. Schuck, Phys. Lett. B{\bf 179}, 313 (1986).
\bibitem{NEB86} G. Nebbia {\it {et al.}}, Phys. Lett. B {\bf 176}, 20 (1986).
\bibitem{HAG88} K. Hagel {\it {et al.}}, Nucl. Phys. A{\bf 486}, 429 (1988).
\bibitem{CHB91} A. Chbihi, L.G. Sobotka, N.G. Nicolis, D.G. Sarantities, D.W. Stracener, Z. Majka, D.C. Hensley, J.R. Beene, and M.L. Halbert, Phys. Rev. C {\bf 43}, 666 (1991).
\bibitem{FOR91} B. Fornal {\it {et al.}}, Phys. Rev. C {\bf 44}, 2588 (1991).
\bibitem{YOS92} K. Yoshida {\it {et al.}}, Phys. Rev. C {\bf 46}, 961 (1992).
\bibitem{SHL91} S. Shlomo and J.B. Natowitz, Phys. Rev. C {\bf 44}, 2878 (1991).
\bibitem{DE98} J.N. De, S. Shlomo, and S.K. Samaddar, Phys. Rev. C {\bf 57}, 1398 (1998).

\end{thebibliography}
\end{document}